\def \SAIT #1 #2 {{\em Mem.\ Soc.\ Astron.\ It.\/} {\bf #1}, #2}
\def \MESS #1 #2 {{\em The Messenger\/} {\bf #1}, #2}
\def \ASTRNACH #1 #2 {{\em Astron. Nach.\/} {\bf #1}, #2}
\def \AAP #1 #2 {{\em Astron. Astrophys.\/} {\bf #1}, #2}
\def \AAL #1 #2 {{\em Astron. Astrophys. Lett.\/} {\bf #1}, L#2}
\def \AAR #1 #2 {{\em Astron. Astrophys. Rev.\/} {\bf #1}, #2}
\def \AAS #1 #2 {{\em Astron. Astrophys. Suppl. Ser.\/} {\bf #1}, #2}
\def \AJ #1 #2 {{\em Astron. J.\/} {\bf #1}, #2}
\def \ANNREV #1 #2 {{\em Ann. Rev. Astron. Astrophys.\/} {\bf #1}, #2}
\def \APJ #1 #2 {{\em Astrophys. J.\/} {\bf #1}, #2}
\def \APJL #1 #2 {{\em Astrophys. J. Lett.\/} {\bf #1}, L#2}
\def \APJS #1 #2 {{\em Astrophys. J. Suppl.\/} {\bf #1}, #2}
\def \APSS #1 #2 {{\em Astrophys. Space Sci.\/} {\bf #1}, #2}
\def \ASR #1 #2 {{\em Adv. Space Res.\/} {\bf #1}, #2}
\def \BAIC #1 #2 {{\em Bull. Astron. Inst. Czechosl.\/} {\bf #1}, #2}
\def \JSQRT #1 #2 {{\em J. Quant. Spectrosc. Radiat. Transfer\/} {\bf #1}, #2}
\def \MN #1 #2 {{\em Mon. Not. R. Astr. Soc.\/} {\bf #1}, #2}
\def \MEM #1 #2 {{\em Mem. R. Astr. Soc.\/} {\bf #1}, #2}
\def \PLR #1 #2 {{\em Phys. Lett. Rev.\/} {\bf #1}, #2}
\def \PASJ #1 #2 {{\em Publ. Astron. Soc. Japan\/} {\bf #1}, #2}
\def \PASP #1 #2 {{\em Publ. Astr. Soc. Pacific\/} {\bf #1}, #2}
\def \NAT #1 #2 {{\em Nature\/} {\bf #1}, #2}

\def\etal{{\it et~al.}}
\def\amin{\ifmmode^{\prime}\else$^{\prime}$\fi}
\def\asec{\ifmmode^{\prime\prime}\else$^{\prime\prime}$\fi}

\def\simgt{\lower.5ex\hbox{$\; \buildrel > \over \sim \;$}}
\def\simlt{\lower.5ex\hbox{$\; \buildrel < \over \sim \;$}}

\newcommand\xte{{\it RXTE}}

\newcommand\asca{{\it ASCA}}

\newcommand\rosat{{\it ROSAT}}

\def\snr{\hbox{N157B}}
\def\psr{\hbox{PSR J0537$-$6910}}
\def\lmcpsr{\hbox{PSR B0540$-$69}}
\def\wg{Wang \& Gotthelf 1998}

\documentstyle{memsait}
\input epsf.sty
\begin{opening}
\title{A 16 MILLISECOND X-RAY PULSAR IN THE CRAB-LIKE SNR N157B: FAST TIMES AT 30 DORADUS} 
\author{E. V. Gotthelf$^1$, W. Zhang$^1$, F. E. Marshall$^1$, J. Middleditch$^2$ \& \\ Q. D. Wang$^3$}
\institute{$^1$NASA/Goddard Space Flight Center, Code 660.2, Greenbelt, MD 20771. \\
$^2$ Los Alamos National Laboratory, MS B256, CIC-19, Los Alamos, NM 87545. \\ $^3$ Dearborn Observatory, Northwestern University,
2131 Sheridan Road, Evanston, IL 60208.
}
\date{} 
\end{opening}

\begin{document}

\oddpagefooter{}{}{} 
\evenpagefooter{}{}{} 
\ 
\bigskip

\begin{abstract}

The supernova remnant \snr\ (30 Dor B, SNR 0539-69.1, NGC 2060),
located in the Tarantula Nebula of the Large Magellanic Cloud, has
long been considered a possible Crab-like remnant. This hypothesis has
been confirmed, quite spectacularly, with the discovery of \psr, the
remarkable 16 ms X-ray pulsar in \snr. \psr\ is the most rapidly
spinning pulsar found to be associated with a supernova remnant.  Here
we report our discovery and summarize the properties of this pulsar and
its supernova remnant.

\end{abstract}

\section{Introduction}

The young Crab-like supernova remnants (SNRs) with their embedded
pulsars provide a critical link connecting the supernova phenomena,
supernova remnants, and the birth of neutron stars.  These rare SNRs
are distinguished by their centrally-filled morphologies and
non-thermal X-ray spectra.  The emission of such SNRs comes
predominantly from a synchrotron nebula powered by the young pulsar
(e.g., Seward 1989). Only four confirmed members of this class were
previously known: the Crab Nebula, the MSH 15--52 nebula, SNR
B0540$-$693, and G11.2--0.3, with its recently discovered 65 ms pulsar
(Torii 1997).

Several candidate Crab-like SNR have been identified. These share
similar morphologies and spectral characteristics but lack an observed
pulsar. Here we report on one of these, N157B (Henize 1956), a SNR
coexisting with a young OB association LH99 in the Large Magellanic
Cloud (LMC). This SNR had been suggested as a Crab-like remnant, or a
``plerion'', because of its centrally peaked morphologies and flat
spectra in both radio and X-ray (Wang \& Gotthelf 1998 and ref. therein). 
Recently, a point-like X-ray
source, possibly the putative pulsar, has been identified using a deep
high resolution \rosat\ image (Wang \& Gotthelf 1998).

In this proceedings we present \psr, the newly discovered ultra-fast
pulsar in \snr\ (Marshall \etal\ 1998) and highlight its
properties. The serendipitous detection of pulsed X-ray emission
provided conclusive confirmation for the Crab-like nature of this
remnant.

\section{Observations and Results}

During a search for pulsed X-ray emission from the recent supernova
SN~1987A, we discovered a highly significant signal at 62 Hz (Marshall
\etal\ 1998).  The signal was found in \xte\ PCA data acquired in 1996 
and located to within 1 degree of the supernova.  The magnitude of the pulsed
flux, however, was too high to be from SN~1987A. An
examination of known X-ray sources within the PCA field-of-view (FOV)
reveals several objects with significant flux in its $2-60$ keV
bandpass. These included the bright black-hole candidate LMC X-1, the
50 ms pulsar \lmcpsr, the R136 star formation region, and the
supernova remnant \snr.

Fortuitously, this LMC field, because of the presence of SN1987A, has
been extensively sampled since 1987 by many X-ray observatories
including \asca. The Gas Imaging Spectrometers (GISs) on-board \asca\
provide arcminute imaging over its $2-10$ keV energy bandpass with
high timing accuracy similar to that of the PCA ($\sim 100 \mu s$).

A pulsar search of the \asca\ data overlapping the region covered by the PCA
FOV unambiguously identified the pulsed signal with \snr, which lies
just $14^{\prime}$ away from SN1987A. This is most vividly
demonstrated by the image of the pulsed emission alone (Fig. 1); this
reveals a single source of significant pulsed emission, an \asca\
point-like source located at the coordinates of the SNR \snr.

\bigskip
\begin{figure}[here]
\epsfxsize=6.0cm
\centerline{\hfill \epsfbox[50 50 500 500]{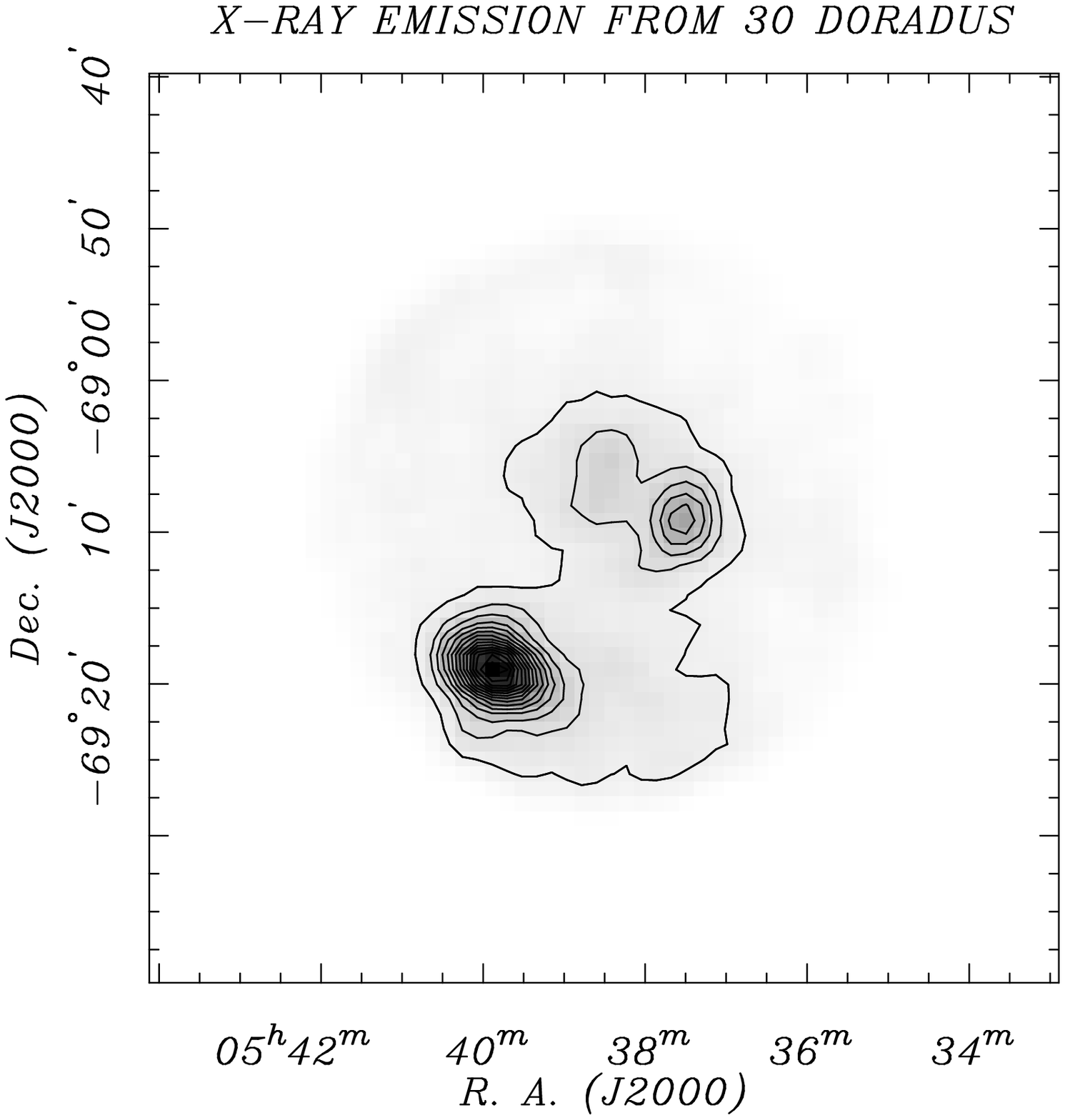} \hspace{1cm} \epsfxsize=6.0cm \epsfbox[50 50 500 500]{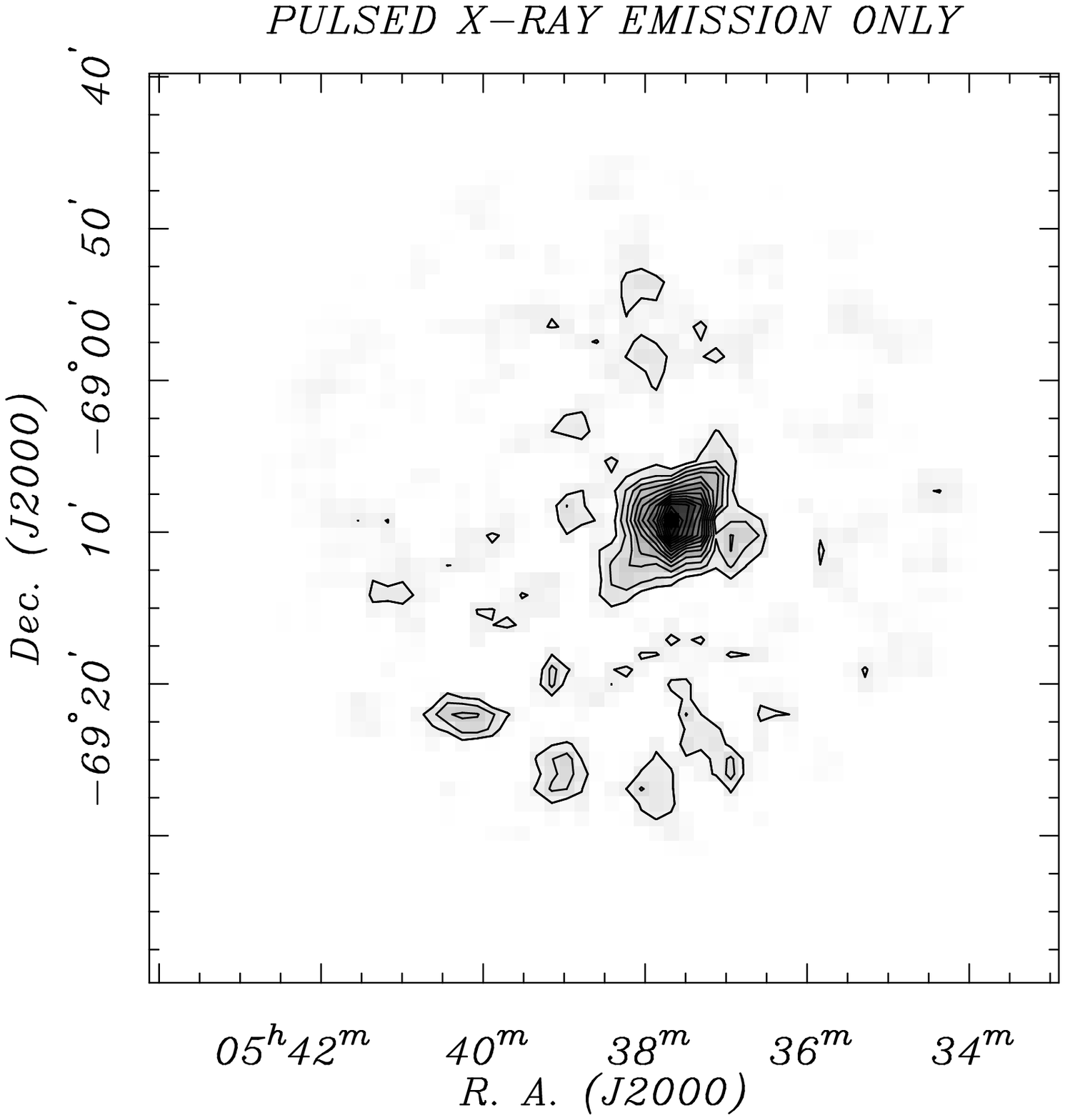} \hfill}
\caption{The 30 Doradus region of the LMC as imaged in X-rays by \asca.
(Left) The broad-band flux corrected, smoothed GIS image reveals the bright 
SNR B0540$-$693, the R136 sources, and SNR N157B. (Right) The same region
after subtracting off the unpulsed image from the pulsed data. The only significant 
pulsed flux is coincident with the X-ray emission associated with N157B.
}
\end{figure}

N157B is a barely resolved, moderately bright \asca\ source in the 30
Doradus star formation region. \snr\ is clearly imaged in the \asca\
GIS as an isolated source, $\sim 10^{\prime}$ southwest of the R136
region, and $\sim 15^{\prime}$ northwest of the \lmcpsr.
The origin of the unpulsed X-rays detected from \snr\ in the energy
band $\ge 2$~keV is most likely a synchrotron nebula powered by the pulsar
via a relativistic wind (Gallant \& Arons 1994 and ref.
therein).  Recent work by Wang \& Gotthelf (1998) has shown that \snr\
contains a bright, elongated, and non-thermal X-ray feature whose
origin is still uncertain. Attached to this feature is a compact
source with a spatial extent of $\sim 7^{\prime\prime}$, which they
claim likely represents the pulsar and its interaction with
the surrounding medium. 

Although the radio morphology is similar to that seen in the X-rays,
searches for a radio pulsar have been unsuccessful. The sensitivity of
these searches, however, may not be well constraining for a Crab-like
object at the distance to the LMC (Kaspi, per. comm.). The remnant has probably
evolved to a stage close to that of a composite SNR.  The elongated
feature is most likely due to relativistic particles transported from
the pulsar to a radio-emitting wind bubble in a collimated outflow; a
substantial fraction of high energy particles from the pulsar lose
their energy radiatively in a bow shock, rather than adiabatically
through diffusion (see Wang \& Gotthelf 1998 for details).

\begin{figure}[here]
\epsfxsize=6.0cm
\centerline{\hfill \epsfbox[50 50 500 500]{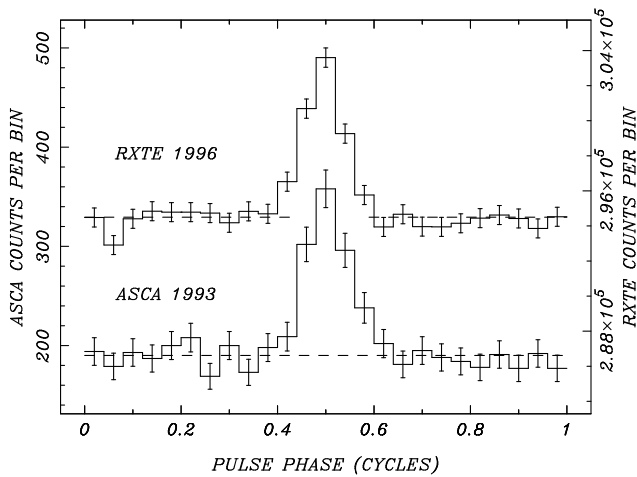} \hfill} 
\caption{The pulse profile of \psr\ in the $2-10$ keV band from the
\xte\ (top) and \asca\ (bottom) observations. The two profiles have
been aligned to place the peak emissions at the 0.5 phase bin. The
relative phases of the two measurements are arbitrary. The \xte\
profile includes 100 ks of data; the \asca\ profile is restricted to
the 37 ks of 64 $\mu \rm{s}$ resolution data.
}
\end{figure}

The pulse profiles for the 1993 \asca\ and the 1996 \xte\ combined data sets
are displayed in Fig 2. They are well characterized by a single,
narrow peaked pulse with an approximately Gaussian shape of FWHM of
$\sim 1.7$ ms (10\% duty cycle).  Its X-ray pulse profile, the
narrowest among the young pulsars, appears unchanged between the two
observations. The pulse width indicates that the size of the emission
region is likely smaller than about half of the light cylinder radius.

From the archival data spanning a 3.5 year baseline we find a period
derivative of ${\dot P} = 5.126 \times 10^{-14} \ {\rm s \ s^{-1}}$.
The inferred $B$-field for a rotationally powered pulsar is $\sim 1
\times 10^{12}$ Gauss, a typical value for the Crab-like pulsars.  The
characteristic spin-down age ($\sim 5000$ years) is much greater than
any of the other Crab-like pulsars, yet similar to the age estimate
for \snr\ based on X-ray measurements of the size and temperature of
the remnant (Wang \& Gotthelf 1998). Chu \etal\ (1992) have placed a
possible upper bound on the age of the remnant of $\sim 2 \times
10^4$~yrs, based on the kinematics of H$\alpha$-emitting gas in the
region.

\begin{figure}[here]
\epsfxsize=6.0cm 
\centerline{\hfill\epsfbox[50 50 500 500]{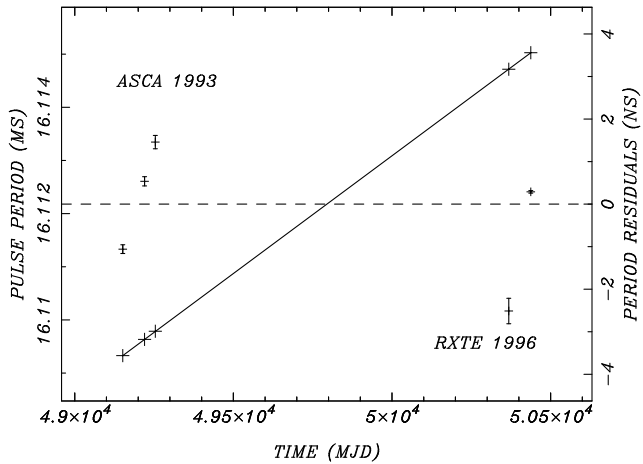}\hfill} 
\caption[h]{The pulse period evolution of \psr\ and its residuals from the best fit model 
(solid line) assuming a linear spin-down trend (see text). The period
measurements are marked by the crosses along the line; the error bars
are smaller then width of the line on this scale (left ordinate). The
period residuals using this model are denoted by the points with error
bars (right ordinate).}
\end{figure}

The X-ray spectrum of \snr, which is dominated by the unpulsed
emission, can be characterized by a power law with an energy slope
$\sim 1.5$ (Wang \& Gotthelf 1998), significantly steeper than those
($\sim 1.0$) of other Crab-like SNRs (e.g., Asaoka \& Koyama
1990). The pulsed emission is also similar to other Crab-like pulsars
with a power law of photon index of $\sim 1.6$. The unabsorbed $2-10$
keV pulsed flux of \psr\ is $\sim 6.7 \pm 0.6 \times 10^{-13} \ {\rm
ergs \ s^{-1} \ cm^{-2}}$.  The corresponding pulsed luminosity is
$\sim 1.7 \times 10^{35} \ {\rm ergs \ s^{-1}} $ (into $4 \pi$)
assuming a distance of 47 kpc (Gould 1995). The derived pulsed
luminosity suggests that $\sim 4 \times 10^{-4}$ of the pulsar's spin
down energy is emitted as pulsed X-rays in the $2-10$ keV band,
consistent with the empirical relation for the magneto-rotation driven 
X-ray emission from previously known pulsars (\"Ogelman 1995)

{PSR J0537$-$6910} rotates twice as fast as the Crab pulsar, and was
likely spinning much more rapidly when it was born, depending on its
assumed braking index and age. Following Kaspi \etal\ (1997), we plot
in Fig. 4 the possible initial spin period vs. age for a range of
braking index. If the older age estimate is correct and if glitches
are not important in the overall period evolution of the pulsar, the
braking index would then have to be unusually small ($\le 1.2$) for an
initial period $\ge 1$ ms.  If the younger value for the age is more
accurate, the initial spin is then a few ms, assuming $n \sim 3$, as
in the case of the magnetic dipole model and for other young pulsars
($n \sim 2.5$). In either case, these values provide important
constraints on neutron star birth spin models.

\begin{figure}[here]
\epsfxsize=8.0cm
\centerline{\epsfbox{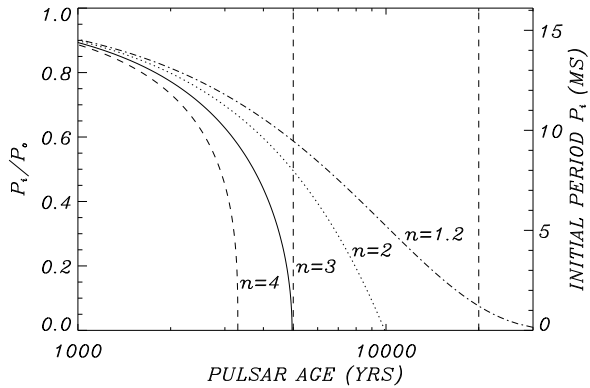} \hfill} 
\caption{Predicted initial period of \psr\ as a function of the age of
the pulsar.  Here we have assumed a power-law deceleration model $\dot
P \propto P^{2-n}$ (e.g., Shapiro \& Teukolsky 1983). The four curves
corresponds to different values of the index: 1.2 (dot-dashed), 2
(dotted), 3 (solid), and 4 (dashed).  The vertical lines indicate a
pulsar age of $5,000$ \& $20,000$ yrs old.  }
\end{figure}

There is compelling evidence that \psr\ has undergone one or more
glitches between the \asca\ and \xte\ epochs (see Fig. 3). The
integrated magnitude of the glitches, $\Delta \Omega / \Omega
\sim 10^{-6}$, is comparable to that observed from the Vela pulsar,
but substantially larger than any of those observed from the Crab
pulsar. More careful analysis of the data is still needed, however, to
quantify the presence of any glitches and to the distinguish them from
timing noise.

\section{Summary of Properties of \psr}

\noindent This pulsar is remarkable in the following ways:

\begin{itemize}
\item 
It is the most rapidly rotating pulsar associated with a SNR, twice as
fast as the Crab pulsar but with a spin-down rate an order of
magnitude slower. 


\item The X-ray pulse profile appears to be the narrowest among the
young pulsars. The Crab pulsar is double-peaked and the profiles of
both \lmcpsr\ and PSR B1509$-$58 are very broad (duty cycle $>
30\%$). The pulse width of the N157B pulsar ($10\%$) indicates that
the size of the emission region is likely smaller than about half of
the light cylinder radius.

\item Both the pulsar's spin-down rate and
the large-scale diffuse emission of the remnant (\wg) suggest
an age of $\sim 5 \times 10^3$~yrs, which is much greater than any of
the other Crab-like pulsars. Thus, N157B has probably evolved to a
stage close to that of a composite SNR. 

\item 
The pulsar shows evidence for glitch(es) between the 3.5 years
separating the \xte\ and \asca\ observations. The integrated magnitude
of the glitches, $\Delta \Omega / \Omega \sim 10^{-6}$, is comparable
to those observed from the Vela pulsar.

\end{itemize}

\begin{itemize}
\item N157B contains a
bright, elongated, and non-thermal X-ray feature whose origin is still
uncertain. Attached to this feature is a compact source, which
most likely represents the pulsar, and possibly its interaction with
surrounding medium. 

\item The X-ray power-law energy slope ($\sim
1.5$) of the N157B SNR is significantly steeper than those ($\sim 1.0$) of
other Crab-like SNRs (e.g., Asaoka \& Koyama 1990), indicating
something unusual may be happening in the remnant.

\item Like the nearby Crab-like \lmcpsr, N157B is in the
Large Magellanic Cloud; we can infer physical parameters of the
remnant with a good distance determination and with relatively little
confusion along the line of sight.

\item N157B is the fifth confirmed Crab-like remnant. It doubles the number of 
known Crab-like remnants/pulsars in the LMC!

\end{itemize}

Detecting the predicted pulsar in N157B confirms its Crab-like nature
and makes it an important new addition to this rare class of objects. 





\end{document}